\documentstyle[12pt]{article}

\newcommand{\beq}{\begin{equation}}
\newcommand{\eeq}{\end{equation}}
\newcommand{\beqa}{\begin{eqnarray}}
\newcommand{\eeqa}{\end{eqnarray}}
\newcommand{\beqar}{\begin{eqnarray*}}
\newcommand{\eeqar}{\end{eqnarray*}}

\newcommand{\Ga}{\Gamma}
\newcommand{\ka}{\kappa}
\newcommand{\inn}{\!\cdot\!}

\renewcommand{\l}{\lambda}

\newcommand{\sig}{\sigma}

\newcommand{\z}{\zeta}

\newcommand{\eg}{{\it e.g.,}\ }
\newcommand{\ie}{{\it i.e.,}\ }
\newcommand{\labell}[1]{\label{#1}} 
\newcommand{\reef}[1]{(\ref{#1})}
\newcommand\prt{\partial}
\newcommand\veps{\varepsilon}
\newcommand\ls{\ell_s}
\newcommand\cF{{\cal F}}
\newcommand\cL{{\cal L}}
\newcommand\cG{{\cal G}}

\newcommand\bz{\bar{z}}
\newcommand\bw{\bar{w}}
\newcommand\hF{\hat{F}}
\newcommand\hA{\hat{A}}
\newcommand\hT{\hat{T}}
\newcommand\htau{\hat{\tau}}
\newcommand\hD{\hat{D}}

\newcommand\ha{\hat{a}}

\newcommand\tG{{\widetilde G}}

\newcommand\tphi{{\tilde \phi}}
\newcommand\tF{{\tilde F}}
\newcommand\tE{{\tilde E}}
\newcommand\tpsi{{\tilde \psi}}
\newcommand\tX{{\tilde X}}
\newcommand\tD{{\tilde D}}
\newcommand\tS{{\tilde S}}

\newcommand\tA{{\widetilde A}}
\newcommand\tT{{\widetilde T}}

\newcommand\tV{{\widetilde V}}
\newcommand\Tr{{\rm Tr}}

\begin{document}

\thispagestyle{empty}
\rightline{\small hep-th/xxxxxxx \hfill IPM/P-2000/013}
\vspace*{1cm}

\begin{center}
{\bf \Large 
Tachyon couplings on non-BPS D-branes\\[.25em]
and Dirac-Born-Infeld action  }
\vspace*{1cm}

{Mohammad R. Garousi\footnote{E-mail:
garousi@yahoo.com}}\\
\vspace*{0.2cm}
{\it Department of Physics, University of Birjand, Birjand, Iran}\\
\vspace*{0.1cm}
and\\
{\it Institute for Studies in Theoretical Physics and Mathematics IPM} \\
{P.O. Box 19395-5746, Tehran, Iran}\\
\vspace*{0.4cm}

\vspace{2cm}
ABSTRACT
\end{center}
By explicit evaluation of certain disk S-matrix elements in the presence
of background B-flux, we 
find  coupling of two open string tachyons to
gauge field, graviton, dilaton or Kalb-Ramond antisymmetric tensor on 
the world-volume of a single non-BPS D$p$-brane. 
We then 
propose an extension of the
abelian Dirac-Born-Infeld action which  naturally
reproduces these couplings in field theory. This action includes  non-linearly
the dynamics of
the tachyon field much like the other  bosonic modes of the non-BPS D$p$-brane.  
On the general grounds of gauge and T-duality transformations and the symmetrized
trace prescription,
we then extend the abelian action  to non-abelian cases.
\vfill
\setcounter{page}{0}
\setcounter{footnote}{0}
\newpage

\section{Introduction} \label{intro}

Recent years have seem dramatic progress in the understanding
of non-perturbative aspects of string theory \cite{excite}.
With these studies has come the realization that solitonic extended
objects, other than just strings, play an essential role.
An important object in these investigations has been Dirichlet
branes \cite{joep}. D-branes are non-perturbative states on which open
string can live, and to which various closed strings including
Ramond-Ramond states can couple.

Type II string theories have two kind of D-branes, BPS D$p$-branes for
$p$ even(odd) \cite{joep}  and non-BPS D$p$-branes for $p$ odd(even) \cite{bergman}
 in
IIA(IIB) theory. The BPS branes are stable solitons which break half
of the space-time supersymmetries and their dynamics are properly described 
in field theory by Dirac-Born-Infeld(DBI) action \cite{bin} (see also \cite{newts})
 and Chern-Simons
action \cite{douglas}. The non-BPS D$p$-branes
on the other hand suffer from open string tachyonic mode whose mass causes
the brane in 
a flat background to be unstable. 
However, there are other terms in the tachyon potential
which  makes it bounded from below. Consequently, the non-BPS branes 
decay
to minimum of the tachyon potential.
It has been  conjectured that at the stationary point of the tachyon 
potential, the negative minimum energy
of the tachyon potential plus the positive energy of the brane tension  
is exactly zero \cite{sen5}.
Hence, the unstable non-BPS branes in flat space-time vacuum should 
decay to the true
vacuum of the theory in which there is no branes. This conjecture 
was studied in \cite{sen7} by explicit
calculation of the tachyon potential using the string field theory framework.

The  dynamics of massless bosonic excitations 
of  non-BPS D$p$-branes are suitablely described by
 the DBI action in field theory. 
This action has been generalized to the supersymmetric form
to include the dynamics of massless 
fermionic modes of the branes as well \cite{sen1}. 
The RR fields of the type II theory 
have no coupling to the non-BPS D-branes through the usual
Chern-Simons action. However, there is a non-vanishing coupling between
the RR field and tachyon on the world-volume of the branes \cite{sen}. The
Chern-Simons action hence was modified in \cite{billo} to incorporate this coupling.
In the present  paper, on the other hand, we are
interested in  generalizing the DBI action to take into account
  the dynamics of the tachyon field.
We study this by explicit
evaluation of some nontrivial disk S-matrix elements in the first quantized
string theory. From these matrix elements we conjecture an extension for
the DBI action which includes the tachyon field as well.

An outline of the paper is as follows. We begin in section 2 by expressing our
conjecture for extension of the DBI action which includes dynamics of the
tachyon field.  Then we
expand this
action around a  background B-flux to produce various couplings involving two tachyons and one gauge field,
graviton, dilaton or Kalb-Ramond antisymmetric tensor. 
In Section 2.1 we transform the above
couplings between commutative fields  
to their non-commutative counterparts. We do this because the
disk S-matrix elements in the presence of the background B-flux with which
we are going to compare our conjectured action in the subsequent section are
corresponding to non-commutative 
open string fields \cite{sw}. 
In Section 3, we evaluate  the S-matrix elements 
and check their  consistency with the  proposed field theory
couplings. In 
Section 4 we extend our proposed action for describing dynamics of a single 
non-BPS D$p$-brane to the case of 
the non-abelian theory of $N$ coincident branes
using the general grounds of the symmetrized trace, and
non-abelian
gauge  and  
invariance under T-duality transformations.
Appendix contains our conventions and some useful comments  
on conformal field theory propagators and
vertex operators used in our calculations.

\section{Abelian action}

The world-volume theory of a single non-BPS D-brane in type II theory 
includes a massless
U(1) vector $A_a$,   a set of massless scalars $X^i$, describing the transverse
oscillations of the brane, a tachyonic state $T$ and their fermionic partners
(see, \eg \cite{sen}). The leading order low-energy action for the massless
fields corresponds to a dimensional reduction of a ten dimensional
U(1) Yang Mills theory. As usual in string theory, there are
higher order $\alpha'=\ls^2$ corrections,
where $\ls$ is the string length scale. As long as derivatives
of the field strengths (and second derivatives of the scalars)
are small compared to $\ls$, then the action takes a Dirac-Born-Infeld
form \cite{bin}. 
To take into account the couplings of the massless open string states
with closed strings, the DBI
 action may be extended naturally to include massless Neveu-Schwarz
closed string fields,\ie  the metric, dilaton and
Kalb-Ramond filed. In this case  one arrives at the 
following world-volume
action:
\beqa
S&=&-T_p \int d^{p+1}\sig\ e^{-\Phi}\sqrt{-{\rm det}(P[G_{ab}+
B_{ab}]+2\pi\alpha'\,F_{ab}}\,\, .\nonumber
\eeqa
Here, $F_{ab}$ is the abelian field
strength of the world-volume ordinary gauge field, while
the metric and antisymmetric tensors are
the pull-backs of the bulk tensors to the D-brane world-volume, \eg
\beqa
P[G_{ab}]&=&G_{\mu\nu}\frac{\prt X^{\mu}}{\prt \sig^a}
\frac{\prt X^{\nu}}{\prt\sig^b}\nonumber\\
&=&G_{ab}+2G_{i(a}\,\prt_{b)}X^i+G_{ij}\prt_aX^i\prt_bX^j
\labell{pull}
\eeqa
where in the second line above we have used that fact that we are employing static
gauge throughout the paper, \ie $\sig^a=X^a$ for world-volume and 
$X^i(\sig^a)$ for
transverse coordinates. 

In order to extend this action to incorporate  dynamics of the tachyonic mode as well, 
we shall
evaluate some disk S-matrix elements in string theory and read from them various 
couplings involving the tachyon field. Our results indicate that the tachyon 
should appear  in the following extension of DBI action:
\beq
S=-T_p \int d^{p+1}\sig\ e^{-\Phi}V(T)\sqrt{-{\rm det}(P[G_{ab}+
B_{ab}]+2\pi\alpha'\,F_{ab}+2\pi\alpha'\prt_a T\prt_b T)}
\labell{biact}
\eeq
where the tachyon potential is $V(T)=1+2\pi\alpha'm^2T^2/2+O(T^4)$ and
the tachyon mass is $m^2=-1/2\alpha'$.
In our conventions the tachyon field is dimensionless. 
The conjecture 
in \cite{sen5} is
that the tachyon potential is zero at the minimum of the potential, \ie
$V(T_0)=0$. Hence, upon tachyon condensation at this point the abelian  action \reef{biact} 
of the non-BPS brane becomes zero. 

Note that  appearance of the tachyon kinetic term 
and  potential in \reef{biact} is similar in form 
to the kinetic term and potential of the 
 transverse scalar fields in the non-abelian
DBI action of $N$ coincident BPS D-branes \cite{myers}. In this case though
the kinetic term  appears in the
pull-back of the metric under the square root and the potential for
scalar fields
multiplies the
square root in the DBI action(see eq.~\reef{nonab} for $T=0$).

We now continue backward,
assuming the above action \reef{biact} 
and check its consistency with  some S-matrix elements.
To have nontrivial check,
we shall evaluate disk  amplitudes describing decay of two tachyons to
dilaton, graviton or Kalb-Ramond antisymmetric tensor on the world-volume
of a single non-BPS D$p$-brane with background
B-flux. The amplitude describing the world-volume 
coupling of two tachyons to gauge field will be evaluated as well. 
Therefore, we begin by   
expanding \reef{biact}  
for fluctuations around the background
$G_{\mu\nu}=\eta_{\mu\nu}$, $B_{\mu\nu}=\cF^{ab}\eta_{a\mu}\eta_{b\nu}$,
$\Phi=0$ and $T=0$ to extract interactions expected from the proposed action
\reef{biact}. The fluctuations
should be normalized as the conventional field theory modes which
appear in the string vertex operators. As a first step, we recall
that the graviton vertex operator corresponds to string frame metric. Hence, one
should transform the Einstein frame metric $G_{\mu\nu}$ to the string frame
metric $g_{\mu\nu}$ via
$G_{\mu\nu}=e^{\Phi/2}g_{\mu\nu}$.
Now with our conventions for string vertex operators (see Appendix), the string mode fluctuations take the
form
\beqa
g_{\mu\nu}&=&\eta_{\mu\nu}+2\kappa h_{\mu\nu}\nonumber\\
\Phi&=&\sqrt{2}\kappa\phi\nonumber\\
B_{\mu\nu}&=&\cF^{ab}\eta_{a\mu}\eta_{b\nu}-2\kappa b_{\mu\nu}
\nonumber\\
T&=&\frac{1}{\sqrt{2\pi\alpha'T_p}}\tau\nonumber\\
A_a&=&\frac{1}{2\pi\alpha'\sqrt{T_p}}a_a\nonumber\\
X^i&=&{1\over\sqrt{T_p}}\l^i\,\, .\nonumber
\eeqa
With these normalizations, the pull back of the Einstein frame metric becomes:
\beqa
P[G_{ab}]&=&\eta_{ab}(1+\frac{\ka}{\sqrt{2}}\phi)+2\ka P[h_{ab}]+
\frac{1}{T_p}(1+\frac{\ka}{\sqrt{2}}\phi)\prt_a\l^i\prt_b\l_i+
\cdots\nonumber
\eeqa
where the dots represents terms with two and more closed string fields.  

Now  it is straightforward,
to expand eq.~\reef{biact} using
\beqar
\sqrt{{\rm det}(M_0+M)}&=&\sqrt{{\rm det}(M_0)}(
1+{1\over2}\Tr(M_0^{-1}M)\nonumber\\
&&-{1\over4}\Tr(M_0^{-1}MM_0^{-1}M)+{1\over8}(\Tr(M_0^{-1}M))^2
+\ldots)
\eeqar
to produce a vast array of interactions. We are  interested in
the interactions quadratic in tachyon, and linear  in massless open and
closed string fluctuations. 
The 
appropriate Lagrangian are: 
\beqa
\cL_{2,0}&=&-c\left(\frac{1}{2}m^2\tau^2+
{1\over2}(V_S)^{ab}\prt_a\tau\prt_b\tau\right)
\labell{int3}\\
\cL_{3,0}&=&-\frac{c}{2\sqrt{T_p}}\left(\frac{1}{2}m^2(V_A)^{ab}f_{ba}\tau^2+\frac{1}{2}V^{ab}f_{ba}
V^{cd}\prt_c\tau\prt_d\tau-V^{ab}f_{bc}V^{cd}\prt_d\tau\prt_a\tau\right)\,\,=\,\, 0
\nonumber\\
\cL_{2,1}&=&-\ka c\left((V^{ab}(h_{ba}-b_{ba})+
\frac{1}{2\sqrt{2}}(\Tr(V)-4)\phi
)
({1\over2}(V_S)^{ab}\prt_a\tau\prt_b\tau+\frac{1}{2}m^2\tau^2)\right.
\nonumber\\
&&\left.\qquad\qquad\qquad\qquad\qquad-V^{ab}(h_{bc}-b_{bc}+\frac{1}{2\sqrt{2}}\phi\eta_{bc})
V^{cd}\prt_d\tau\prt_a\tau
\right)\,\, .
\nonumber
\eeqa
where we have dropped  some total derivative terms in $\cL_{3,0}$. In the 
above Lagrangian, 
$f_{ab}=\prt_a a_b-\prt_b a_a$ and 
\beqa
c\,\,\equiv\,\,\sqrt{-{\rm det}(\eta_{ab}+\cF_{ab})}&\,\,,\,\,&
V^{ab}\,\,\equiv\,\, \left( (\eta+\cF)^{-1}\right)^{ab} \,\,, 
\labell{vmatrix}
\eeqa
and $V_S(V_A)$ is symmetric(antisymmetric) part of the V matrix above.
It is important to note that the antisymmetric matrix $V_A$ appears  in
total derivative terms in the Lagrangian $\cL_{3,0}$ 
which involves only open string fields.

In the case that the background B-flux is zero, the coupling of Kalb-Ramond
field to tachyon in the second 
line of $\cL_{2,1}$ vanishes, and the  
graviton and dilaton couplings reduce to the natural coupling of these fields
to the kinetic term of the tachyon field, \ie $e^{-\Phi}G^{ab}\prt_aT\prt_bT$.
In that way, there is no nontrivial coupling that confirms the conjectured
action \reef{biact} is valid or not. So we continue our discussion for
non-vanishing background B-flux.

The open string fields appearing in the DBI action \reef{biact}  or \reef{int3} are ordinary commutative
fields. Whereas,  open string vertex operators in string theory 
with background B-flux
correspond to non-commutative fields \cite{sw}. Hence, in order to compare the
couplings in \reef{int3} with corresponding S-matrix elements, one has to transform the commutative
fields in \reef{int3} to their non-commutative variables.

\subsection{Change of variables}

In \cite{sw}  differential equation for transforming commutative gauge field to its
non-commutative counterpart
was found  to be
\beqa
\delta \hA_a(\theta)\!\!\!&=\!\!\!&\frac{1}{4}\delta\theta^{cd}\left(\hA_c*\hF_{ad}+\hF_{ad}*A_c-\hA_c*\prt_d\hA_a
-\prt_d\hA_a*A_c\right)\labell{delf}\\
\delta\hF_{ab}(\theta)\!\!\!&=\!\!\!&\frac{1}{4}\delta\theta^{cd}\left(
2\hF_{ac}*\hF_{bd}+2\hF_{bd}*\hF_{ac}
-\hA_c*(\hD_d\hF_{ab}+\prt_d\hF_{ab})
-(\hD_d\hF_{ab}+\prt_d\hF_{ab})*\hA_c\right)
\nonumber
\eeqa
where the gauge field strength and $*$ product were defined to be
\beqa
\hF_{ab}&=&\prt_a\hA_b-\prt_b\hA_a-i\hA_a*\hA_b+i\hA_b*\hA_a\nonumber\\
&=&\prt_a\hA_b-\prt_b\hA_a-i[\hA_a,\hA_b]_M\nonumber\\
\hat{f}(x)*\hat{g}(x)&=&e^{\frac{i}{2}\theta_{ab}\prt^a_{y}\prt^b_{z}}
\hat{f}(y)\hat{g}(z)|_{y=z=x}\,\, .\nonumber
\eeqa
These differential equations can be integrated perturbatively to find a relation
between ordinary fields appearing in \reef{int3} and non-commutative fields corresponding
to open string vertex operators. The result for abelian case that we are interested in
is \cite{mine}:
\beqa
A_a&=&\hA_a-\frac{1}{2}\theta^{cd}\left(\hA_c*'\hF_{ad}-\hA_c*'\prt_d\hA_a\right)+O(\hA^3)
\nonumber\\
F_{ab}&=&\hF_{ab}-\theta^{cd}\left(\hF_{ac}*'\hF_{bd}-\hA_c*'\prt_d\hF_{ab}\right)
+O(\hA^3)
\labell{fhf}
\eeqa
where the commutative multiplication $*'$ operates as
\beqa
\hat{f}(x)*'\hat{g}(x)&=&\frac{\sin(\frac{1}{2}\theta_{ab}\prt^a_{y}\prt^b_{z})}
{\frac{1}{2}\theta_{ab}\prt^a_{y}\prt^b_{z}}\hat{f}(y)\hat{g}(z)|_{y=z=x}\,\, .\nonumber
\eeqa
In \cite{mine} we verified by explicit calculation of S-matrix elements of
one massless closed and two open string states that the  transformation
\reef{fhf} is 
exactly reproduced by perturbative  string theory. 
Appropriate transformations for scalar fields such as the tachyon can be
read from \reef{fhf}, \ie
\beqa
T&=&\hT+\theta^{cd}\hA_c*'\prt_d \hT+\cdots
\labell{TT}\\
\prt_a T&=&\prt_a\hT-i[\hA_a,\hT]_M+\theta^{cd}\left(\hF_{ca}*'\prt_b\hT+
\hA_c*'\prt_a\prt_d\hT\right)+\cdots \nonumber
\eeqa
where dots represent terms which involve  more than two open string fields. 
They produce couplings between  more than three fields upon replacing 
them into \reef{int3} 
in which we are not interested. 

The differential equation \reef{delf}  expresses infinitesimal variation 
 of linear field 
, \eg $\delta \hA_a$, in terms of infinitesimal variation of the 
non-commutative parameter, \ie $\delta\theta^{cd}$.
Upon integration \reef{fhf}, this transforms the linear commutative gauge 
field in terms of
nonlinear combination of non-commutative fields. 
However, we are interested in transforming
quadratic  combinations of commutative fields appearing in \reef{int3} 
in terms of non-commutative fields.
Such a transformation, in principle, might  be found from a  differential 
equation alike \reef{delf} that
 expresses infinitesimal  variation of the quadratic  fields 
in terms of infinitesimal  variation of
non-commutative parameter. Upon integration, that would produced the desired
 transformation.
In that way, one would find that not only the fields transform 
as in \reef{TT}
but the multiplication rule between  fields also  undergo
appropriate transformation.  
We are not going to find such a differential equations here. Instead, we simply
note that  the transformation for multiplication rule between two open
string fields can be conjectured from the right hand side of eq.~\reef{fhf} to be
\beqa
fg|_{\theta=0}&\longrightarrow&f*'g|_{\theta\ne 0}
\labell{fg}
\eeqa
for  $f$ and $g$ being any arbitrary open string fields. This transformation rule
was confirmed in \cite{mine} by explicit evaluation of S-matrix elements of
one massless closed and two open string states.

Now with the help of equation \reef{TT} and \reef{fg}, one can transform 
the commutative Lagrangian \reef{int3} to non-commutative counterparts.
In doing so, one should first
using \reef{fg} replace ordinary multiplication of two tachyons by 
the $*'$ multiplication. Then, using \reef{TT} the ordinary tachyon fields
 should be transformed
to their non-commutative counterparts. The results are
\beqa
{\hat{\cL}_{2,0}}&=&-c\left(\frac{1}{2}m^2\htau\htau+\frac{1}{2}(V_S)^{ab}\prt_a\htau\prt_b\htau\right)+
\frac{ic}{4\pi\sqrt{T_p}}(V_S)^{ab}\prt_a\htau[\ha_b,\htau]_M+\cdots
\labell{hatl}\\
{\hat{\cL}_{2,1}}&=&-\ka c\left((V^{ab}(h_{ba}-b_{ba})+
\frac{1}{2\sqrt{2}}(\Tr(V)-4)\phi
)
({1\over2}(V_S)^{ab}\prt_a\htau*'\prt_b\htau+\frac{1}{2}m^2\htau*'\htau)\right.
\nonumber\\
&&\left.\qquad\qquad\qquad\qquad-V^{ab}(h_{bc}-b_{bc}+\frac{\phi\eta_{bc}}{2\sqrt{2}})
V^{cd}\prt_d\htau*'\prt_a\htau\right)+\cdots\labell{l20}
\eeqa
where ellipsis represent terms which have more than three fields. 
Here we have dropped some
total derivative terms which appeared in $\hat{\cL}_{2,0}$ and also replaced $*'$ 
between two non-commutative tachyons in \reef{hatl} with ordinary
multiplication rule because the deference is some total derivative terms. 
In the eq.~\reef{l20} on the other hand,  the difference between $*'$ 
and ordinary multiplication rules is
not just a total derivative terms. 
Note that upon inserting
the transformation \reef{TT} into \reef{int3}, the antisymmetric
matrix $(V_A)^{ab}$  appears  in eq.~\reef{hatl} only in the $*$ product
terms\footnote{Note that our conventions set $\theta^{ab}=4\pi (V_A)^{ab}$.}.

It is interesting to note that the symmetric part of the $(\eta+\cF)^{-1}$
matrix, \ie $V_S$, appears as the metric in \reef{hatl} and the antisymmetric
part appears in the definition of $*$ product in the Moyal bracket. This is
consistent with the conclusion reached in \cite{sw}. In our discussion, however,
the symmetric part $V_S$ appears naturally as a result of expanding the ordinary
DBI action around the background B-flux, 
and the antisymmetric part $V_A$ appears as a result of transforming
commutative fields to non-commutative variables. In the Lagrangian \reef{l20}, on the other hand,
which involves  open and closed string fields, both symmetric and antisymmetric
matrices  appear in its different coupling terms.

\section{Scattering Calculations}

The couplings in \reef{hatl} and \reef{l20}
should be reproduced by disk S-matrix elements of string theory 
if the proposed action \reef{biact}
is going to be valid. In this section, we work  
at the string theory side and evaluate  these 
couplings using the conformal
field theory technique. We begin with 
the evaluation of the coupling of two tachyons
to gauge or scalar fields.

\subsection{Open-Open-Open couplings}

In the world-sheet conformal field theory framework, 
the coupling of two tachyons to
a gauge or scalar field
is described properly by 
the correlation of their corresponding vertex operators
inserted at the boundary of the disk world-sheet, that is 
\beqa
A&\sim&(\z_3\inn\cG)_{\mu}\int dx_1dx_2dx_3 \langle:V_{-1}(2k_1\inn V^T,x_1)::
V_{-1}(2k_2\inn V^T,x_2)::V^{\mu}_0(2k_3\inn V^T,x_3):\rangle
\nonumber
\eeqa
where the  vertex operators are given in the Appendix. 
Using the world-sheet
conformal field theory technique, 
it is not difficult to perform the correlators 
above and
show that the integrand is invariant under $SL(2,R)$. Gauging 
this symmetry by fixing the positions of the vertices at arbitrary points, one
finds $A(\tau_1,\tau_2,\l_3)=0$ and
\beqa
A(\tau_1,\tau_2,a_3)&=&\frac{c\sin(\pi l)}{\pi\sqrt{T_p}}(k_1\inn V_S\inn\z_3)
\labell{tta}
\eeqa
where we have defined $l\equiv -2k_1\inn V^T\inn \cF\inn V\inn k_2
=2k_1\inn V_A\inn k_2$. 
We have also normalized the amplitude by the appropriate coupling 
factor $-c/2\pi\sqrt{T_p}$. 
The $\sin(\pi l)$ factor above arises basically from two different phase factors
corresponding to 
two distinct cyclic orderings of the vertex operators. Each phase factor
stems
from the second 
term of the world-sheet propagator \reef{pro2}.
Using the fact that our conventions set $
\theta^{ab}=4\pi V^{ab}_A$, it is not difficult to verify that the S-matrix element
\reef{tta} is exactly reproduced by the second term in \reef{hatl}. At the same time,
vanishing of $A(\tau_1,\tau_2,\l_3)$ is consistent with \reef{hatl}.

\subsection{Closed-Open-Open amplitudes} \label{quadi}

The amplitudes
describing decay of two open string tachyons to  one massless 
closed string NSNS mode 
is given by the following correlation:
\[
A\sim(\veps_3\inn D)_{\mu\nu}\int dx_1dx_2d^2z\langle:V_0(2k_1\inn V^T,x_1)::
V_0(2k_2\inn V^T,x_2)::
V_{-1}^{\mu}(p_3,z_3)::V_{-1}^{\nu}(p_3\inn D,\bz_3):\rangle
\]
where the closed string vertex operator inserted at the middle  and open 
string vertex operators  at the boundary of the disk world-sheet.
Explicit form of the vertex operators in terms of world-sheet
fields  are given in the Appendix.
Here again using appropriate world-sheet propagators, 
one can evaluate the 
correlations above and show that the integrand is $SL(2,R)$ invariant.
We refer the reader to Refs.~\cite{ours,aki,scatd} for the details of the
calculations.
Gauging the $SL(2,R)$ symmetry by fixing  $z_3=i$ and $x_1=\infty$, one arrives at
\beqa
A&\sim&2^{-2s-2}\int dx\left( (2s+1)\Tr(\veps_3\inn D)-
\frac{8ik_2\inn V^T\inn\veps_3\inn D\inn V\inn k_1 
}{x-i}+\frac{8ik_1\inn V^T\inn\veps_3\inn D\inn V\inn k_2
}{x+i}\right)\nonumber\\
&&\qquad\qquad\qquad\times (x-i)^{s-l}(x+i)^{s+l}\nonumber
\eeqa
where the integral is taken from $-\infty$ to $+\infty$, and 
$s=-p_3\inn V_S\inn p_3=-1/2-2k_1\inn V_S\inn k_2$.
This integral is doable and the result is
\beqa
A&=&-\frac{i\ka c}{2}\left(a_1(s+l)-a_2(s-l)
\right)\frac{\Ga(-2s)}{\Ga(1-s-l)\Ga(1-s+l)}
\labell{Ansnst}
\eeqa
where $a_1$ and $a_2$ are two kinematic factors depending only on the 
space time momenta and the closed string polarization tensor
\beqa
a_1&=&-4k_2\inn V^T\inn\veps_3\inn D\inn V\inn k_1\nonumber\\
a_2&=&(s+l)\Tr(\veps_3\inn D)+
4k_1\inn V^T\inn\veps_3\inn D\inn V\inn k_2\,\, .\nonumber
\eeqa
We have also normalized the amplitude \reef{Ansnst} at 
this point by the coupling factor $-i\ka c/{2\pi}$. 
As a check of our calculations, we have inserted the dilaton polarization \reef{vdilaton}
into \reef{Ansnst} and found that the result is independent of the auxiliary vector $\ell^{\mu}$.
The amplitude \reef{Ansnst} has the  pole structure 
at  $m_{open}^2=n/{\alpha'}$\footnote{We explicitly restore $\alpha'$ here.
Otherwise our conventions set $\alpha'=2$}. This does not have tachyon pole
which is consistent with the fact that coupling of three tachyons is zero. In fact
due to the world-sheet fermions in the tachyon vertex operator, coupling of any odd number
of 
tachyons is zero. Hence, the world-volume of the non-BPS D$p$-branes has a
$Z_2$ symmetry under which the tachyon changes sign.

\subsubsection{Massless poles}

Given the general form of the string amplitude in eq.~\reef{Ansnst}, one can
expand this amplitude as an infinite sum of 
terms reflecting the infinite tower
of open string states that propagate on the world-Volume of D-brane.
In the  domain where  $s\longrightarrow 0$, 
 the first term of the expansion
representing the exchange of massless string states dominate. In this case
the scattering amplitude \reef{Ansnst} reduces to
\beqa
A&=&\frac{i\ka c\sin(\pi l)}{4\pi s}(a_1+a_2)
+\cdots
\labell{Atau}
\eeqa
where dots represent contact terms and the infinite 
series of massive poles.
Making the appropriate explicit choices of polarizations, we find
\beqa
A_s(\tau_1,\tau_2,\phi_3)&=&\frac{i\ka c\sin(\pi l)}
{8\pi\sqrt{2}s}(\frac{l}{2}(\Tr(D)+2)
-4k_1\inn V\inn V\inn k_2)
+1\leftrightarrow 2
\labell{llt}
\\
A_s(\tau_1,\tau_2, h_3)&=&\frac{i\ka c\sin(\pi l)}{4\pi s}(\frac{l}{2}
\Tr(\veps_3\inn D)
-4k_1\inn V\inn \veps_3^T\inn V\inn k_2)+1\leftrightarrow 2\,\, .
\nonumber
\eeqa
Here $h_3$ stands for both graviton and Kalb-Ramond antisymmetric tensors.
In writing explicitly the above massless poles,  
one finds some terms which 
are proportional to $s$ as well. We will
add these terms which have no contribution 
to the massless poles of field theory to 
the contact terms in
\reef{contact}. The amplitudes \reef{llt} should be reproduced in $s$-channel 
of field theory. 
They 
can be evaluated in field theory as
\beqa
A'_s(\tau_1,\tau_2,\phi_3)&=&(\tV_{\phi_3 a})^a(\tG_a)_{ab}(\tV_{a\tau_1\tau_2})^b\nonumber\\
A'_s(\tau_1,\tau_2,h_3)&=&(\tV_{h_3 a})^a(\tG_a)_{ab}(\tV_{a\tau_1\tau_2})^b
\labell{llt'}
\eeqa
where the propagator and the vertices can be read from expansion of
\reef{biact} in terms of non-commutative fields. They are 
\beqa
(\tG_a)^{ab}&=&\frac{i}{c}\frac{(V_S^{-1})^{ab}}{s}\nonumber\\
(\tV_{\phi_3 a})^a&=&\frac{\sqrt{T_p}\ka c}{2\sqrt{2}}\left(
\frac{1}{2}(\Tr(D)+2)p_3\inn V_A^a-p_3\inn V\inn V^a+V^a\inn V\inn p_3\right)
\nonumber\\
(\tV_{h_3 a})^a&=&\sqrt{T_p}\ka c\left(
\frac{1}{2}\Tr(\veps_3\inn D)p_3\inn V_A^a-p_3\inn V\veps_3^T\inn V^a+V^a\inn\veps_3^T V\inn p_3\right)
\nonumber\\
(\tV_{a\tau_1\tau_2})^a&=&\frac{c\sin(\pi l)}{2\pi\sqrt{T_p}}
k_1\inn V_S^a+1\leftrightarrow 2\,\, .\nonumber
\eeqa
In writing the above propagator, we have used the covariant
gauge $V_S^{ab}\prt_a \hA_b=0$. 
Replacing above propagator and vertices into \reef{llt'}, one finds exactly
the string massless poles \reef{llt}.

\subsubsection{Contact terms}

Having examined in detail the massless poles of string amplitude, we now
extract the low energy contact terms of the string amplitude \reef{Ansnst}. 
Expanding the gamma
function appearing in this amplitude for $s\longrightarrow 0$, one will find
\beqa
A&=&\frac{i\ka c}{2}\left((\frac{a_1+a_2}{2\pi})
\frac{\sin(\pi l)}{s}+(\frac{a_1-a_2}{2})
\frac{\sin(\pi l)}{\pi l}\right.\nonumber\\
&&\left.+(a_1+a_2)\frac{\sin(\pi l)}{\pi l}\,
\sum_{n=1}^{\infty}\zeta(2n+1)l^{(2n+1)}+k^2O(s,l)\right)
\,\, .\nonumber
\eeqa
The factor $\sin(\pi l)/(\pi l)$ appears for all the contact terms which
is consistent with the transformation of 
 multiplication rule in \reef{fg}. 
The second term of the first line above is the contact term 
with minimum number of momentum in which we are interested, that is,
\beqa
A_c&\equiv&\frac{i\ka c\sin(\pi l)}{4\pi l}(a_1-a_2)
\labell{contact}
\eeqa
Inserting 
appropriate polarization (see Appendix) and adding the residue
contact terms of the massless poles \reef{Atau}, one finds
\beqa
A_c(\tau_1,\tau_2,\phi_3)&=&\frac{i\ka c\sin(\pi l)}{8\pi\sqrt{2}l}\left(
-\frac{s}{2}(\Tr(D)+2)
-4k_1\inn V\inn V\inn k_2\right)+1\leftrightarrow 2
\labell{allphi}\\
A_c(\tau_1,\tau_2,h_3)&=&\frac{i\ka c\sin(\pi l)}{4\pi l}\left(
-\frac{s}{2}\Tr(\veps_3\inn D) 
-4k_1\inn V\inn\veps_3^T\inn V\inn 
k_2
\right)
+1\leftrightarrow 2\,\, . \nonumber
\eeqa
where again $h_3$ stands for both graviton and Kalb-Ramond antisymmetric tensors.
These contact terms are reproduced exactly by the Lagrangian in \reef{l20}. The
first terms in \reef{allphi} by the terms in the 
first line of \reef{l20} and the second terms in \reef{allphi} by the terms
in the second
line of \reef{l20}. Note that, while the first terms in \reef{allphi} can
be reproduced in field theory by an action in which the tachyon kinetic
term appears linearly like the one proposed in \cite{sen1}, 
 the second terms in \reef{allphi} can be reproduced
only if  the tachyon kinetic term appears non-linearly in 
the determinant under the square root in the BDI action, \ie eq.~\reef{biact}. 
This ends our illustration of
consistency between disk S-matrix elements and  the proposed 
action  \reef{biact}.

\section{Non-abelian action}
In this section we extend the proposed action \reef{biact} for a non-BPS
D$p$-brane to the case of N coincident non-BPS D$p$-brane where the world-volume theory
involves a U(N) gauge theory. Our guiding principle in constructing such a non-abelian
action is that the action should be consistent with the familiar rules of 
T-duality. This guideline   has been  recently employed by Myers 
\cite{myers} to construct
non-abelian DBI action. In this way, one should start with non-abelian action
for D9-branes and then use some sort of T-duality transformations to convert the D9-brane action
to non-abelian action for D$p$-branes. Therefore, we begin by  extending
the abelian action
\reef{biact} to non-abelian action for non-BPS D9-branes. 
In this case there is no scalar field corresponding
to the transverse direction of D9-branes. Hence, the non-abelian 
action may be constructed from the corresponding abelian case  by
simply extending  the derivative of open string fields to  its 
non-abelian covariant derivative \cite{hull}, and a trace over
the U(N) representations
\cite{tsytlin} (see also
\cite{newts}), that is,
\beq
S=-T_9 \int d^{10}\sig\ \Tr\left(e^{-\Phi}V(T)\sqrt{-{\rm det}(G_{\mu\nu}+
B_{\mu\nu}+2\pi\alpha'\,F_{\mu\nu}+2\pi\alpha'D_{\mu} TD_{\nu} T)}\right)
\labell{biact9}
\eeq
where $G_{\mu\nu}$, $B_{\mu\nu}$ and $F_{\mu\nu}$ are the metric, antisymmetric tensor and 
the non-abelian gauge field strength, respectively,
and 
\beqa
D_{\mu}T&=&\frac{\prt T}{\prt\sig^{\mu}}-i[A_{\mu},T]\,\, .\nonumber
\eeqa
This action is still incomplete without a precise prescription
for how the gauge trace should be implemented. We expect that a prescription
similar to that given  for Born-Infeld action \cite{tsytlin} should also be given 
here. That is, the gauge trace should be completely symmetric between
all non-abelian expression of the form $F_{\mu\nu}$, $D_{\mu}T$ and
individual $T$ appearing in the tachyon potential $V(T)$. 

Now we generalize \reef{biact9} to the action 
appropriate
for  non-BPS D$p$-branes for any  $p$. To this end,
we apply familiar T-duality transformations rules to the non-abelian
D9-brane action \reef{biact9}. T-duality transformations in $i=p+1,\cdots,9$ directions
of the D9-brane world-volume converts the D9-brane to D$p$-brane, 
the gauge fields in those direction to
$\tA_i=X^i/2\pi\alpha'$ and leaves  unchanged the tachyon, \ie $\tT=T$. The new
scalar fields $X^i$ are now transverse coordinates of the new D$p$-brane. Under this
transformation, the covariant derivative of tachyon becomes
\beqa
\tD_i\tT&=&\frac{\prt T}{\prt \sig^i}-\frac{i}{2\pi\alpha'}[X^i,T]\,\,\, .
\nonumber
\eeqa
Using the fact that we are  always working in the static gauge,
the first term on the right hand side becomes  zero because of 
the assumption implicit in the
T-duality transformations that all fields must be independent of the coordinates
$\sig^i$. Now adding this transformation to the T-duality transformation of massless
fields (see \eg \cite{myers}), one has complete list of T-duality transformations
for the fields appearing in \reef{biact9}; 
\beqa
\tE_{ab}\,=\,E_{ab}-E_{ai}E^{ij}E_{jb}&,&\tE_{ai}\,=\,E_{aj}E^{ji}\nonumber\\
\tE_{ij}\,=\,E^{ij}&,&E_{ia}\,=\,-E^{ij}E_{ja}\nonumber\\
e^{2\tphi}\,=\,e^{2\phi}{\rm det}(E^{ij})&,&\tD_i\tT\,=\,-\frac{i}{2\pi\alpha'}[X^i,T]\labell{tdual}\\
\tF_{ab}\,=\,F_{ab}&,&\tF_{ai}\,=\,\frac{1}{2\pi\alpha'}D_aX^i\nonumber\\
\tF_{ij}\,=\,-\frac{i}{(2\pi\alpha')^2}[X^i,X^j]&,&\tF_{ia}\,=\,-\frac{1}{2\pi\alpha'}D_aX^i\nonumber
\eeqa
where we have defined $E_{\mu\nu}\equiv G_{\mu\nu}+B_{\mu\nu}$. Here $E^{ij}$
denotes the inverse of $E_{ij}$, \ie $E^{ik}E_{kj}=\delta^i{}_j$.
Under above T-duality transformations the determinant in  \reef{biact9} becomes
\beqa
\tD\,=\,{\rm det}\left(\matrix{\!\!\!\!&\tE_{ab}
+2\pi\alpha'F_{ab}+2\pi\alpha'D_aTD_bT&\tE_{aj}+D_aX^j-iD_aT[X^j,T]\cr\nonumber\\ \nonumber
\!\!\!\!&\tE_{ib}-D_bX^i-i[X^i,T]D_bT&\tE_{ij}-\frac{i}{2\pi\alpha'}[X^i,X^j]
-\frac{1}{2\pi\alpha'}[X^i,T][X^j,T]\cr}\right)
\nonumber
\eeqa
Manipulating the matrix inside the determinant, one finds
\beqa
\tD&=&{\rm det}\left(P[E_{ab}+E_{ai}(Q^{-1}-
\delta)^{ij}E_{jb}]+2\pi\alpha'F_{ab}
+T_{ab}\right){\rm det}(Q^i{}_j){\rm det}(E^{ij})\labell{tD}
\eeqa
where now the definition of the pull-back above is the  extension of
\reef{pull}  in which 
ordinary derivative is replaced by its non-abelian covariant derivative. Here 
the matrices  $Q^i{}_j$ and $T_{ab}$ are defined to be
\beqa
Q^i{}_j&=&\delta^i{}_j-\frac{i}{2\pi\alpha'}[X^i,X^k]E_{kj}-\frac{1}{2\pi\alpha'}[X^i,T][X^k,T]E_{kj}\nonumber\\
T_{ab}&=&2\pi\alpha'D_aTD_bT-D_aT[X^i,T](Q^{-1})_{ij}[X^j,T]D_bT\labell{qt}\\
&&-iE_{ai}(Q^{-1})^i{}_j[X^j,T]D_bT-iD_aT[X^i,T](Q^{-1})_i{}^jE_{jb}\nonumber\\
&&-iD_aX^i(Q^{-1})_{ij}[X^j,T]D_bT-iD_aT[X^i,T](Q^{-1})_{ij}D_bX^j
\nonumber
\eeqa
In equations \reef{tD} and \reef{qt}, indices are raised and lowered by $E^{ij}$
and $E_{ij}$, respectively. Now replacing \reef{tD} into T-dual of \reef{biact9} and using the transformation for 
dilaton field \reef{tdual}, one finds the final T-dual action 
\beqa
\tS&=&-T_p\int d^{p+1}\sig\labell{nonab}\\
&&\times\Tr\left(e^{-\Phi}V(T)V'(T,X^i)
\sqrt{-{\rm det}(P[E_{ab}+E_{ai}(Q^{-1}-\delta)^{ij}
E_{jb}]+2\pi\alpha'F_{ab}+T_{ab})}\right)\nonumber
\eeqa 
where we have defined $V'(T,X^i)=\sqrt{{\rm det}(Q^i{}_j)}$. This potential term is
one for abelian case.
If the tachyon field is set to zero, this action would get 
to the result of non-abelian action for N coincident BPS D$p$-branes \cite{myers}. 
In this case, the prescription for the gauge trace is studied in \cite{myers}.
The trace is completely symmetric between $F_{ab}$, $D_a X^i$, $i[X^i,X^j]$
and individual $X^i$. The latter non-abelian field stems from
non-abelian Taylor expansion of the closed string fields that appear in
the DBI action \cite{gm}. Natural extension of this prescription for the trace  
in the action \reef{nonab} is that the trace should be completely 
symmetric between all non-abelian expressions of the form $F_{ab}$,
$D_aX^i$, $i[X^i,X^j]$, $X^i$, $D_aT$, $i[X^i,T]$ and individual $T$
of the tachyon potential.


{\bf Acknowledgments}

I would like to acknowledge useful conversation with R.C. Myers. 
This work was supported by University of Birjand and IPM.

\newpage
\appendix
\section{ Perturbative string theory with background field}

In perturbative superstring theories, to study scattering amplitude of 
some external string states 
  in conformal
field theory frame,
one usually evaluate correlation function of their corresponding vertex
operators with
use of some  standard conformal field theory 
propagators \cite{pkllsw}.
In trivial flat background one uses an  appropriate 
{\it linear} $\sigma$-model to
derive the propagators and define the vertex operators.
 In
nontrivial
D-brane background the vertex operator remain unchanged while the standard
propagators need some modification. Alternatively, one may use a doubling
trick to convert the propagators to standard form and give the modification
to the vertex operators \cite{ours}. 
In this appendix we would like to consider a D-brane with constant gauge field 
strength / or antisymmetric Kalb-Ramond field in  all directions of the D-brane.
The modifications arising from the appropriate {\it linear} $\sigma$-model
 appear in the following boundary conditions \cite{leigh}\footnote{
Our notation and conventions follow  those established in \cite{ours}.
So we are working on the upper-half plane 
with boundary at $y=0$ which means $\prt_y$ is
normal derivative and $\prt_x$ is tangent derivative.
And our index conventions are that lowercase Greek
indices take values in the entire ten-dimensional
space-time, \eg $\mu,\nu=0,1,\ldots,9$; early Latin indices take values
in the world-volume, \eg $a,b,c=0,1,\ldots,p$; and middle Latin indices
take values in the transverse space, \eg $i,j =p+1,\ldots,8,9$.
Finally, our conventions
set $\ls^2=\alpha'=2$.}: 
\beqa
\prt_y X^a-i\cF^a{}_b\prt_x X^b\,\,=\,\,0&{\rm for }&a,b\,=0,1,\cdots p 
\nonumber\\
X^i\,\,=\,\,0&{\rm for}&i\,=p+1,\cdots 9
\labell{mixboundary}
\eeqa
where $\cF_{ab}$  are the  constant background fields, 
and these equations are imposed at $y=0$. 
The world-volume (orthogonal subspace) indices
are raised and lowered by $\eta^{ab}(N^{ij})$ 
and $\eta_{ab}(N_{ij})$, respectively.
Now we have to understand the modification of the conformal field theory
propagators arising from these mixed boundary conditions. To this end consider
the following general expression for propagator of $X^{\mu}(z,\bz)$ fields:
\beqa
<X^{\mu}(z,\bz)\,X^{\nu}(w,\bw)>
&=&-\eta^{\mu\nu}\log(z-w)-\eta^{\mu\nu}\log(\bz-\bw) \nonumber\\
&&-D^{\mu\nu}\log(z-\bw)-D^{\nu\mu}(\bz-w)
\labell{pro1}
\eeqa
where $D^{\mu\nu}$ is a constant matrix. 
To find this matrix, we impose
the boundary condition \reef{mixboundary} on the propagator \reef{pro1}, which 
yields
\beqa
\eta^{ab}-D^{ba}-\cF^{ab}-\cF^a{}_c D^{bc}&=&0\labell{D1}
\eeqa
for the world-volume directions, $D^{ij}=-N^{ij}$ for the orthogonal directions, and 
$D^{ia}=0$ otherwise.
Now equation \reef{D1} can be solved for $D^{ab}$, that is
\beqa
D_{ab}&=&2(\eta-\cF)^{(-1)}_{ab}-\eta_{ab}\labell{D2}\\
&=&2V_{ba}-\eta_{ab}
\eeqa
where matrix $V$ is the dual metric  that appears 
in the expansion of DBI action \reef{vmatrix}.
Note that the $D^{\mu\nu}$ is orthogonal matrix, 
\ie $D^{\mu}{}_{\alpha}D^{\nu\alpha}=\eta^{\mu\nu}$.  

Using two dimensional equation of motion, one can write 
the world-sheet fields in terms of right- and left-moving components. In terms of
these chiral fields,
closed NSNS and open NS vertex operators are
\beqa
V^{\rm NSNS}&=&:V_n^{\rm NS}(X(z),\psi(z),\phi(z),p):
:V_m^{\rm NS}(\tX(\bz),\tpsi(\bz),\tphi(\bz),p):
\nonumber\\
V^{\rm NS}&=&:V_n^{\rm NS}(X(x)+\tX(x),\psi(x)+\tpsi(x),\phi(x)+\tphi(x),k):\nonumber
\eeqa
where $\psi^{\mu}$ is super partner of world-sheet field $X^{\mu}$ and $\phi$
is world-sheet superghost field. The indices  $n,m$ refer to the superghost charge of vertex operators,
and $p$ and $k$ are closed and open string momentum, respectively.
In order to work with only right-moving fields, we use the following doubling
trick:
\beqa
\tX^{\mu}(\bz)\longrightarrow D^{\mu}{}_{\nu}X^{\nu}(\bz) &\,\,\,\,\,
\tpsi^{\mu}(\bz)\longrightarrow D^{\mu}{}_{\nu}\psi^{\nu}(\bz)&\,\,\,\,\,
\tphi(\bz)\longrightarrow \phi (\bz)\,\, .
\labell{trick}
\eeqa
These replacements in effect extend the right-moving 
fields to the entire complex plane and
shift modification arising from mixed boundary condition
from propagators to vertex operators. Under these replacement, world-sheet
propagator between all right-moving fields take the 
standard form \cite{mrg} except
the following boundary propagator:
\beqa
<X^{\mu}(x_1)\,X^{\nu}(x_2)>&=&-\eta^{\mu\nu}\log(x_1-x_2)+
\frac{i\pi}{2}\cF^{\mu\nu}\Theta(x_1-x_2)
\labell{pro2}
\eeqa
where $\Theta(x_1-x_2)=1(-1)$ if $x_1>x_2(x_1<x_2)$. Note that the orthogonal
property
of the $D$ matrix is an important ingredient for writing the propagators in the standard form. The vertex operators
under transformation \reef{trick} becomes
\beqa
V^{\rm NSNS}&=&:V_n^{\rm NS}(X(z),\psi(z),\phi(z),p):
:V_m^{\rm NS}(D\inn X(\bz),D\inn \psi(\bz),\phi(\bz),p):
\nonumber\\
V^{\rm NS}&=&:V_n^{\rm NS}(X(x)+D\inn X(x),\psi(x)+D\inn\psi(x),2\phi(x),k):\,\, .
\nonumber
\eeqa
The vertex operator for  massless NSNS and NS
states and open string tachyon are
\beqa
V^{\rm NSNS}&=&(\veps\inn D)_{\mu\nu}:V_n^{\mu}(p,z):
:V_m^{\nu}(p\inn D,\bz):\nonumber\\
V^{\rm NS}&=&(\z\inn\cG)_{\mu}:V_n^{\mu}(2k\inn V^T,x):\nonumber\\
V^{\tau}&=&:V_n(2k\inn V^T,x):
\eeqa
where $\cG^{ab}=(\eta^{ab}+D^{ab})/2=V^{ba}$ for gauge field, 
$\cG^{ij}=(\eta^{ij}-D^{ij})/2=N^{ij}$
for scalar field and $\cG^{ai}=0$ otherwise.
The open string  vertex operators in $(0)$ and $(-1)$ pictures are
\beqa
V^{\mu}_0(k,x)&=&\left(\prt X^{\mu}(x)+ik\inn\psi(x)\,
\psi^{\mu}(x)\right)e^{i k\cdot X(x)}\nonumber\\
V^{\mu}_{-1}(k,x)&=&e^{-\phi(x)}\psi^{\mu}(x)e^{i k\cdot X(x)}\nonumber\\
V_0(k,x)&=&ik\inn\psi(x)e^{ik\cdot X(x)}\nonumber\\
V_{-1}(k,x)&=&e^{-\phi(x)}e^{ik\cdot X(x)}\,\, .
\nonumber
\eeqa
The physical conditions for the massless open string and tachyon are
\beqa
{\rm massless}:&&\qquad\qquad k\inn V_S\inn k=0\,\,\,\,\,,\,\,\,\,\, k\inn V_S\inn\z=0\nonumber\\
{\rm tachyon}:&&\qquad\qquad k\inn V_S\inn k=\frac{1}{4}\nonumber
\eeqa
and for massless closed string are $p^2=0$ and $p_{\mu}\veps^{\mu\nu}=0$ where
$\veps$ is the closed string polarization which is 
traceless and symmetric(antisymmetric)
for graviton(Kalb-Ramond) and 
\beq
\veps^{\mu\nu}=\frac{1}{\sqrt{8}}(\eta^{\mu\nu}-\ell^{\mu}p^{\nu}
-\ell^{\nu}p^{\mu})\,\,\,\,\,,\,\,\,\,\, \ell\inn p=1
\labell{vdilaton}
\eeq
for the dilaton. Using the fact that $D^{\mu\nu}$ is orthogonal
matrix, one finds the following identities:
\[
\cG\inn\cG^T=\cG^S\,\,\,,\,\,\,(D\inn\cG^T)^{ab}=\cG^{ab}\,\,\,,\,\,\,
(D\inn\cG^T)^{ij}=-N^{ij}
\]
where the $\cG^S$ is symmetric part of the $\cG$ matrix.


\newpage
\begin{thebibliography}{99}

\bibitem
{excite}{A. Sen, ``An Introduction to Nonperturbative String Theory,''
 hep-th/9802051;\\
C. Vafa, ``Lectures on Strings and Dualities,''
 hep-th/9702201;\\
J. Polchinski, Rev. Mod. Phys. {\bf 68} (1996) 1245
[hep-th/9607050];\\
M.J. Duff, Int. J. Mod. Phys. {\bf A11} (1996) 5623
[hep-th/9608117];\\
J.H. Schwarz, Nucl. Phys. Proc. Suppl. {\bf 55B} (1997) 1
[hep-th/9607201].}

\bibitem
{joep}{J. Polchinski, ``TASI Lectures on D-branes,''
 hep-th/9611050;\\
J. Polchinski, S. Chaudhuri and C.V. Johnson, ``Notes on D-branes,''
hep-th/9602052;\\
W. Taylor, ``Lectures on D-branes, Gauge Theory and M(atrices),''
hep-th/9801182.}

\bibitem
{bergman}{O. Bergman and M.R. Gaberdiel, Phys. Lett. {\bf B441} (1998)133
[hep-th/9806155];\\
A. Sen, JHEP {\bf 9809} (1998) 023 [hep-th/9808141];\\
A. Sen, JHEP {\bf 9810} (1998) 021 [hep-th/9809111];\\
A. Sen, JHEP {\bf 9812} (1998) 021 [hep-th/9812031];\\
P. Horava, Adv. Theor. Math. Phys. {\bf 2} (1999) 1373 [hep-th/9812135].}

\bibitem
{bin}{R.G. Leigh, Mod. Phys. Lett. {\bf A4} (1989) 2767;\\
C.G. Callan, C. Lovelace, C.R. Nappi and S.A. Yost, Nucl. Phys. {\bf B308} (1988) 221;\\
A. Abouelsaood, C.G. Callen, C.R. Nappi and S.A. Yost, Nucl. Phys. {\bf B280} (1987) 599.}

\bibitem
{newts}{A.A. Tseytlin, ``Born-Infeld Action, Supersymmetry And String Theory,''
hep-th/9908105.}

\bibitem
{douglas}{M.R. Douglas, ``Branes within Branes,'' hep-th/9512077;\\
M. Li, Nucl. Phys. {\bf B460} (1996) 351 [hep-th/9510161];\\
M. Green, J.A. Harvey and G. Moore, Class. Quant. Grav. {\bf 14} (1997) 47
[hep-th/9605033].}


\bibitem
{sen5}{A. Sen, JHEP {\bf 9808} (1998) 010 [hep-th/9805019];\\
A. Recknagel and V. Schomerus, Nucl. Phys. {\bf B545} (1999) 233 [hep-th/9811237];\\
C.G. Callan, I.R. Klebanov, A.W. Ludwig and J.M. Maldacena,
Nucl. Phys. {\bf B422} (1994) 417 [hep-th/9402113];\\
J. Polchinski and L. Thorlacius, Phys. Rev. {\bf D50} (1994) 622 [hep-th/9404008];\\
A. Sen, Int. J. Mod. Phys. {\bf A14} (1999) 4061 [hep-th/9902105].}

\bibitem
{sen7}{A. Sen, JHEP {\bf 9912} (1999) 027  [hep-th/9911116];\\
A. Sen and B. Zwiebach, ``Tachyon Condensation in String Field Theory,''
hep-th/9912249;\\
V.A. Kostelecky and Samuel, Phys. Lett. {\bf B207} (1988)169;\\
W. Taylor, ``D-brane effective field theory from string field theory,''
hep-th/0001201;\\
N. Moeller and W. Taylor, ``Level truncation and the tachyon in open bosonic
string theory,'' hep-th/0002237;\\
J.A. Harvey and P. Kraus, ``D-Branes as Unstable Lumps in Bosonic Open String 
Field Theory,'' hep-th/0002117;\\
N. Berkovits, ``The Tachyon Potential in Open Neveu-Schwarz String Field
Theory,'' hep-th/0001084;\\
N. Berkovits, A. Sen and B. Zwiebach, ``Tachyon Condensation in Superstring Field Theory,''
hep-th/0002211.}

\bibitem
{sen1}{A. Sen, JHEP {\bf 9910} (1999) 008 
[hep-th/9909062].}

\bibitem
{sen}{A. Sen, ``Non-BPS States and Branes in String Theory,'' 
hep-th/9904207.}

\bibitem
{billo}{M. Billo, B. Craps and F. Roose, JHEP {\bf 9906} (1999) 033
[hep-th/9905157].}

\bibitem
{sw}{N. Seiberg and E. Witten, JHEP {\bf  9909 }(1999) 032 [ hep-th/9908142].}





\bibitem
{myers}{R.C. Myers, JHEP {\bf 9912} (1999) 022 [hep-th/9910053].}

\bibitem
{mine}{M.R. Garousi, ``Non-commutative World-volume interactions on D-branes
and Dirac-Born-Infeld action,'' hep-th/9909214.}





\bibitem
{ours}{M.R. Garousi and R.C. Myers,
{ Nucl. Phys.} {\bf B475} (1996) 193 [hep-th/9603194].}



\bibitem
{aki}{A. Hashimoto and I.R. Klebanov, 
{Phys. Lett.} {\bf B381} (1996) 437 [hep-th/9604065].}

\bibitem
{scatd}{For a review, see: A. Hashimoto and I.R. Klebanov, Nucl. Phys. Proc.
Suppl. {\bf 55B} (1997) 118 [hep-th/9611214].}



\bibitem
{hull}{H. Dorn, Nucl. Phys. {\bf B494} (1997) 105 [hep-th/9612120];\\
C.M. Hull, JHEP {\bf 9810} (1998) 011
[hep-th/9711179].}

\bibitem
{tsytlin}{A.A. Tseytlin, Nucl. Phys. {\bf B501} (1997) 41 [hep-th/9701125].}



\bibitem
{gm}{M.R. Garousi and R.C. Myers, Nucl. Phys. {\bf B542} (1999) 73 [hep-th/9809100].}

\bibitem
{pkllsw}{M.E. Peskin, ``Introduction to string and superstring theory II, in
>From the planck scale to the weak scale: toward a theory of the universe,''
Proceedings of TASI's86, H.Haber ed., World Scientific Publishing 1987;\\
V.A. Kostelecky, O. Lechtenfeld, W. Lerche, S. Samuel and S. Watamura,
Nucl. Phys. {\bf B288} (1987) 173.}

\bibitem
{leigh}{J.Dai, R.G. Leigh and J.Polchinski, Mod. Phys. Lett. {\bf A4} (1989) 2073.}

\bibitem
{mrg}{M.R. Garousi, JHEP {\bf 9812} (1998) 008 [hep-th/9805078].}

\end{thebibliography}
\end{document}